\documentclass[onecolumn,amsmath, amssymb, nobibnotes, aps, prf, floatfix, showpacs]{revtex4-2}

\usepackage{graphicx,wasysym,color,booktabs}

\newcommand{\env}[1]{\texttt{#1}}
\newcommand\T{\rule{0pt}{3ex}} 
\newcommand\B{\rule[-1.5ex]{0pt}{0pt}} 

\usepackage{epstopdf}
\usepackage{graphicx}
\usepackage{xcolor}
\usepackage[latin1]{}
\usepackage{tabularx}
\usepackage{placeins}
\usepackage[separate-uncertainty=true,multi-part-units=single]{siunitx}
\usepackage{hyperref}

\usepackage{soul}

\begin{document}

\title{Effects of viscosity on liquid structures produced by in-air microfluidics}

\author{David Baumgartner}
\email[]{david.baumgartner@tugraz.at}
\author{Günter Brenn}
\author{Carole Planchette}
\affiliation{Institute of Fluid Mechanics and Heat Transfer, Graz University of Technology, A-8010 Graz, Austria}
\date{\today}

\begin{abstract}
This study experimentally investigates the effect of viscosity on the outcomes of collisions between a regular stream of droplets and a continuous liquid jet. A broad variation of liquid viscosity of both the drop and the jet liquid is considered, keeping other material properties unchanged. To do so, only two liquid types were used: aqueous glycerol solutions for the drop and different types of silicone oil for the jet liquid. Combining these liquids, the viscosity ratio $\lambda=\mu_{drop}/\mu_{jet}$ was varied between 0.25 and 3.50. The collision outcomes were classified in the form of regime maps leading to four main regimes: \textit{drops-in-jet}, \textit{fragmented drops-in-jet}, \textit{encapsulated drops}, and \textit{mixed fragmentation}. We demonstrate that, depending on the drop and jet viscosity, not all four regimes can be observed in the domain probed by our experiments. The experiments reveal that the jet viscosity mainly affects the transition between \textit{drops-in-jet} and \textit{encapsulated drops}, which is shifted towards higher drop spacing for more viscous jets. The drop viscosity leaves the previous transition unchanged, but modifies the threshold of the drop fragmentation within the continuous jet. We develop a model that quantifies how the drop viscosity affects its extension, which is at first order fixing its shape during recoil and is therefore determining its stability against pinch-off.
\end{abstract}

\begin{description}
\item[PACS numbers]
\end{description}

\maketitle

\section{Introduction}
\label{sec:introduction}

The large number of recent scientific publications dedicated to encapsulation shows the increasing need for reliable, precise and scalable technologies. This demand is mainly motivated by the biomedical and  pharmaceutical industries, which strive to deliver actives as efficiently and safely as possible \cite{ref:Rosen2005, ref:Kumari2010,  ref:Kearney2013, ref:Yoo2011}. The development of cell culture and tissue engineering requires, beyond the necessity of cell feeding and harvesting, a mean, to manipulate and assemble the cells, which can be achieved by their regular and controlled encapsulation into a matrix \cite{ref:Bhatia2005, ref:Nicodemus2008, ref:Wang2014, ref:Steele2014}. The need of encapsulation is also rising in less demanding applications such as in the production of cosmetic, food-products, agricultural inputs, and in depollution tasks \cite{ref:Yeo2004, ref:Brandenberger1998, ref:Haeberle2008, ref:Serp2000, ref:Martins2014, ref:Augustin2009}.

To tackle these challenges, several methods have been proposed. For the production of well controlled spherical capsules, the technology of choice is microfluidics. Indeed, since researchers have been using this toolbox, many micro-droplet based applications emerged, including chemical micro-reactors, multiple emulsions and cell capsules \cite{ref:Zhao2013, ref:Chabert2008, ref:Koester2008, ref:Riche2016}. Yet, while present in the scientific community since decades, microfluidics has barely made it to industries. Beside the need for precise chips requiring appropriate design and manufacture, the risk of clogging remains, the main issue which considerably limits scale-up possibilities \cite{ref:Chiu2017, ref:Dressaire2017, ref:Su2006}. Regarding the production of fibers, which are especially desirable for medical and biomedical applications \cite{ref:Sanchez2016, ref:Hu2014, ref:Enizi2018}, the state of the art relies on coaxial or emulsion electrospinning. The former, however, enables only the production of core-shell structures, and the latter does not offer the control on the size and position of the inclusions \cite{ref:Moghe2008, ref:Yarin2011, ref:Hu2006}.

The previously mentioned drawbacks of these existing technologies call for innovative approaches. Inspired by the important knowledge about drop impacts, which include drop impacting onto a wall \cite{ref:Yonemoto2017, ref:Wildeman2016, ref:Eggers2010, ref:Clanet2004, ref:Yarin1995, ref:Laan2014}, a thin liquid film \cite{ref:Josserand2003, ref:Wang2000, ref:Kittel2018}, a liquid bath \cite{ref:Lhuissier2013, ref:Ray2015} or another drop \cite{ref:Moqaddam2016, ref:Planchette2017, ref:Roisman2012, ref:Ashgriz1990, ref:Roisman2004}, the so called in-air-microfluidics, has recently been proposed \cite{ref:Planchette2018, ref:Kamperman2018}. This promising approach consists in solidifying the liquid microstructures resulting from the collision in air of drop streams and jets \cite{ref:Visser2018}. Binary drop collisions involving two or three drops of one or more liquids \cite{ref:Planchette2012, ref:Hinterbichler2015, ref:Planchette2017}, and the collision of a stream of drops with a continuous jet, count to this rather new encapsulation method. The structures produced by the drop-jet collisions enable to form both spherical capsules and regular fibers containing periodic encapsulation of monodisperse spheres. The collisions taking place in air, it suppresses the need for the additional carrying liquid phase which must be used in microfluidics. Most importantly, the absence of channels eliminates the critical risks of clogging. Finally, the alignment requirements are much more moderate than for drop-drop collisions. 

Beside these obvious advantages, in-air-microfluidics remains to date largely unexplored, and its potential and limits must still be described and understood. Indeed, not much happened since the pioneering work of Chen et al., who used water for both the drops and the jet \cite{ref:Chen2006}. In that study, several behaviours were identified, which were named with increasing inertia as bouncing, coalescence, segmenting, separation and splashing. The next study was performed by Planchette et al., who used immiscible liquids \cite{ref:Planchette2018}. The outcomes were classified according to the fragmentation of the drops or the jet, both or none providing fragmented drops in jet, capsules, mixed fragmentation and drops in jet. The fragmentation of the jet was attributed to a capillary instability, while the one of the drop was associated to an excess of its kinetic energy, which leads to its excessive deformation and thus its fragmentation. The proof of concept regarding the solidification of the produced structures was provided about the same time by Visser et al. \cite{ref:Visser2018}. More recently, the effects of liquid wettability and miscibility have been investigated \cite{ref:Baumgartner2020}. It was shown that replacing total wetting by partial one, or exchanging immiscible liquids against miscible ones, do not significantly modify the outcomes, as long as the surface tension of the jet remains lower than the one of the drops. Questions about the role of the liquid viscosities have not been addressed yet. Our experimental study aims to fill this gap and bring the knowledge of in-air microfluidics one step forward. To do so, we investigate six different liquid combinations consisting of two liquid types, namely three aqueous glycerol solutions for the drops and four silicone oils for the jet liquid.  These liquids provide total wetting of the jet on the drop, promoting their encapsulation, and differ only by their viscosity. First, a wide range of collision parameters are screened for each combination, and the corresponding outcomes are classified in four regimes, following the analysis by Planchette et al. \cite{ref:Planchette2018}. Regime maps are built to evidence the effects of viscosity on these regimes and their occurrence. The results are further interpreted by focusing on the capillary fragmentation of the jet and on the inertial fragmentation of the drop. For the latter, the description of the drop extension and recoil is enabled by the use of two cameras and aliasing stroboscopic illumination. 

The paper is organized as follows. The experimental set-up and problem description are first introduced. The experimental results are then presented, and the shifts in regime boundaries are discussed. The paper ends with the conclusions.

\section{Material and experimental methods}
\label{sec:materialandmethods}
\subsection{Experimental set-up and problem description}
\label{sec:set-up}
\begin{figure}
\centering 
\includegraphics[width=17.8cm]{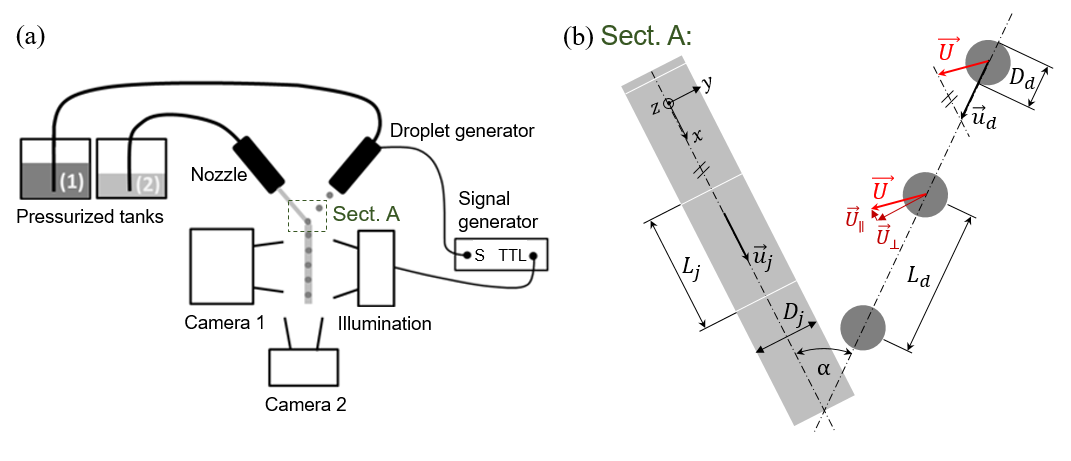}
\caption{(a) Experimental set-up for the drop/jet collision experiments. (b) Geometric and kinetic parameters of the collisions. Figure adapted from \cite{ref:Baumgartner2020}.}
\label{fig:setup}
\end{figure}

The present study focuses on head-on collisions of a regular stream of monodisperse droplets and a continuous immiscible liquid jet. These collisions are realized by adjusting the trajectory of the droplets and the jet into the same plane to avoid off-plane eccentricity. A sketch of the set-up used to generate this kind of collisions is shown in figure  \ref{fig:setup}(a). Two pressurized and independent tanks supply the liquids for the drops and the jet, which are produced with a droplet generator \cite{ref:Brenn1996} and a nozzle, respectively. Micro traverses enable the accurate adjustment of their trajectories. The drops and jet diameters range between $D_d=205\pm25\,\mathrm{\mu m}$ and $D_j=290\pm20\,\mathrm{\mu m}$, respectively. Here, as well as in the rest of this manuscript, the subscript $d$ refers to the drop parameters and the drop liquid properties, while $j$ is used for the jet and the jet liquid properties. The drop generator and the illumination system (stroboscopic LED) are connected to a signal generator in order to supply both devices with the same frequency ($8000\,\mathrm{Hz}<f_d<26000\,\mathrm{Hz}$). This allows the recording of frozen collision pictures.  The imaging of one collision is then performed with two cameras providing orthogonal (Camera 1) and front (Camera 2) views for all experiments. The drops are dyed and the jet remains transparent, providing a strong contrast to easily distinguish them.
All needed collision parameters are extracted from collision pictures using the public-domain software ImageJ (\href{https://imagej.nih.gov/ij/}{https://imagej.nih.gov/ij/}) shown in figure \ref{fig:setup}(b). These parameters include the drop diameter $D_d$ and the jet diameter $D_j$, the spatial period of the drops $L_d$ ($300-1100\,\mu$m) and of the jet $L_j$ ($300-800\,\mu$m), the collision angle $\alpha$ ($15^\circ-60^\circ$), and the velocities of the drop $\vec{u}_d$ ($4-15$ ms$^{-1}$) and the jet $\vec{u}_j$ ($3-15$ ms$^{-1}$). The {magnification is larger than in our previous study \cite{ref:Baumgartner2020} with not more than 6 $\mu m$ per pixel providing uncertainty below 3$\%$ for all measured diameters and velocities}. Further detailed information about the methods applied to obtain these parameters is shown elsewhere \cite{ref:Baumgartner2020}. The relative velocity $\vec{U}$ ($2-10$ ms$^{-1}$) between drop and jet is defined as the impact velocity and is calculated as $\vec{U}=\vec{u}_d-\vec{u}_j$. Note that in this study the component of the relative velocity parallel to the jet trajectory, ${U}_\parallel={u}_d \cos(\alpha)-{u}_j$, is set to zero {with even stricter conditions than in \cite{ref:Baumgartner2020} (${U}_\parallel<0.04{U}$ and even ${U}_\parallel<0.01{U}$ close to the transitions instead of ${U}_\parallel<0.1{U}$)}. As a consequence, the relative velocity $\vec{U}$ corresponds to the component perpendicular to the jet trajectory {whose norm} ${U}_\perp={u}_d \sin(\alpha)$ {is independent from $L_j/D_j=u_j/(f_d D_j)$}. This {adjustment} can be realized by {varying} $\vec{u}_d$, $\vec{u}_j$ and $\alpha$ {to obtain $u_d cos(\alpha)=u_j$, which is the condition ensuring} head-on collisions.

For this work focusing on the viscosity role, we further introduce the viscosity ratio $\lambda=\mu_d/\mu_j$. It relates the dynamic viscosity of the drop liquid $\mu_d$ to the dynamic viscosity of the jet liquid $\mu_j$ and lies between 0.25 and 3.50.  
\FloatBarrier
\subsection{Liquids}
\label{sec:liquids}
We use three aqueous glycerol solutions as the drop liquids and four silicone oils as the jet liquids. These two liquids are immiscible and provide total wetting of the jet on the drop. The interfacial tension between the drop liquids and the jet liquids can be specified as $\sigma_{dj}=32\pm3\,mNm^{-1}$ \cite{ref:Peters2013,ref:Robinson1970,ref:Dong2017}. The values for the surface tension $\sigma$,  the density $\rho$, the dynamic viscosity $\mu$ and {the range of Ohnesorge number $Oh=\mu/\sqrt{\rho \sigma D}$ of all liquids used are shown in table \ref{tab:liquids}. $Oh$ is calculated using either the drop or the jet diameter and the corresponding liquid properties.} The density is measured by weighing an exact volume of 100 ml, the viscosity is determined with a glass capillary viscometer, and the surface tension is measured with the pendant drop method. To obtain aqueous solutions of different viscosities for the drops, the mass fractions of glycerol ($\geq98\%$, Carl Roth GmbH, Germany) in water are 12$\%$, 50$\%$ and 68$\%$ in case of G1, G5 and G20, respectively. The viscosity of the jet liquid is varied by using different silicone oils (SO3, SO5, and SO20 are pure liquids supplied by Carl Roth GmbH, Germany). SO1 is a mixture of SO3 and SO0.65 (supplied by IMCD South East Europe GmbH, Austria) with a mass ratio of 70$\%$:30$\%$. In addition, the drop liquid is dyed with Indigotin 85 (E 132, BASF, Germany) at a concentration of 1g/l. The dye is added to the aqueous glycerol solutions before the liquid properties are measured. 
\begin{table}[h]
\caption{Liquid properties. All measurements were carried out at an ambient temperature of $T_{amb}=23\pm1^{\circ}$C}
\begin{ruledtabular}
\begin{tabular}{ccccc}
\T
Abbreviation & Density & Dynamic viscosity & Surface tension & Ohnesorge number  \\
\B  & $\rho\,(kg \cdot m^{-3})$ & $\mu\,(mPa \cdot s)$ & $\sigma\,(mN \cdot m^{-1})$ & $Oh (-)$\\
\cline{1-5}
\T  G1 & 1024$\pm$5 & 1.33$\pm$0.05 & 70$\pm$1 & 0.011$\pm$0.001\\
G5 & 1118$\pm$5 & 5.05$\pm$0.10 & 68$\pm$2 & 0.041$\pm$0.004\\
G20 & 1169$\pm$5 & 17.14$\pm$0.05 & 67.5$\pm$1 & 0.137$\pm$0.007 \\
SO1 &  846$\pm$5 & 1.45$\pm$0.03 & 17$\pm$1 & 0.023$\pm$0.001\\
SO3 &  887$\pm$5 & 2.73$\pm$0.05 & 18.5$\pm$0.5 & 0.034$\pm$0.001\\
SO5 &  915$\pm$10 & 5.10$\pm$0.05 & 19.5$\pm$0.5 & 0.071$\pm$0.004\\
\B SO20 &  949$\pm$5 & 19.15$\pm$0.05 & 20.5$\pm$0.5 & 0.263$\pm$0.002\\
\end{tabular}
\end{ruledtabular}
\label{tab:liquids}
\end{table}
\FloatBarrier
\section{Results: Regime maps}
\label{sec:regimemaps}
\begin{figure}
\centering 
\includegraphics[width=0.90\textwidth]{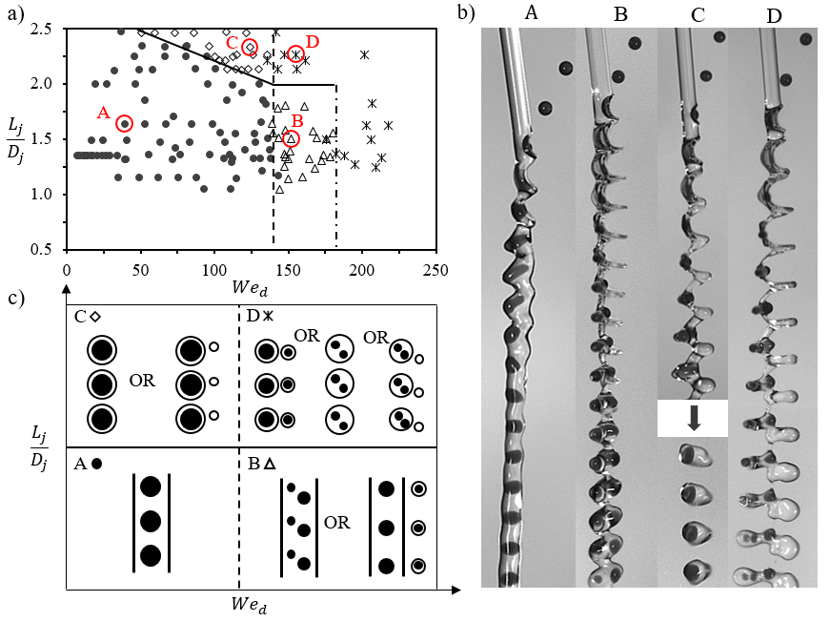}
\caption{(a) Collision outcomes of the reference case G5/SO5 classified in the form of a regime map with $We_d$ and $L_j/D_j$ as the scaling parameters. (b) Recorded collision pictures (front view - camera 2) of the structures produced by drop-jet collisions. A, B, C and D correspond to the data marked in red in (a). (c) Schematic illustration of the structures. Adapted from \cite{ref:Baumgartner2020}.
}
\label{fig:referencecase}
\end{figure}

The first set of experiments using G5 as the drop liquid and SO5 as the jet liquid is defined as the reference case with a viscosity ratio of $\lambda=\mu_d/\mu_j\approx1$. It establishes the starting point for investigating viscosity effects. The results are represented in figure \ref{fig:referencecase}(a) in the form of a regime map, similar to the ones already introduced in \cite{ref:Planchette2018} and \cite{ref:Baumgartner2020}. Collision pictures recorded in front view (camera 2) are shown in figure \ref{fig:referencecase}(b). For clarity, a schematic illustration is added in figure \ref{fig:referencecase}(c). Further details about the observed regimes and the reference case can be found in \cite{ref:Baumgartner2020}. For consistency, we briefly recall the four regimes: 
\begin{itemize}
\item  (A) \textit{drops-in-jet}: The drops are totally engulfed by the jet, which remains continuous. This regime is marked with filled, but differently coloured circles in all regime maps.
\item  (B) \textit{fragmented drops-in-jet}: In this case, the drops fragment, but the jet remains continuous. Often all drop fragments remain inside the continuous jet, but sometimes parts of them may be expelled. This regime is marked with black empty triangles. 
\item  (C) \textit{encapsulated drops}: The droplets do not fragment, but the jet breaks up. The result is a regular stream of droplets, which are encapsulated with the jet liquid. This regime is marked with black empty diamonds.  
\item  (D) \textit{mixed fragmentation}: Here, both the drops and the jet fragment. This regime is marked with black crosses.
\end{itemize} 
The regime maps used for the classification of the collision outcomes is based on two scaling parameters. The first scaling parameter $L_j/D_j$, a purely geometric one, is used to predict the jet break-up. The fragmentation mechanism can be primarily attributed to a capillary-driven instability, similar to the one of Plateau Rayleigh \cite{ref:Planchette2018}. The geometric term $L_j$ describes the distance between two consecutive impacting droplets (see figure \ref{fig:setup}(b)), which is normalized by the diameter of the jet $D_j$. By exceeding a critical value of $L_j/D_j$, the jet breaks into a regular stream of encapsulated droplets  {\cite{ref:Planchette2018, ref:Baumgartner2020, ref:Baumgartner2019, ref:Planchette2017ILASS}}. The critical value of the reference case G5/SO5 is approximately 2, as indicated by the solid line in figure \ref{fig:referencecase}(a). 
 {Note that our previous study \cite{ref:Baumgartner2020} associates the transition between \textit{drops-in-jet} and \textit{encapsulated drops} to a unique critical value of $L_j/D_j$, whereas the present work reveals a moderate influence of inertia on this critical value. Several reasons can explain the slight deviations observed in the data. First and most likely, the  differing points of \cite{ref:Baumgartner2020} were obtained with $U_{\parallel} \approx 0.1 U$ instead of  $U_{\parallel}<0.01$, which is known to cause additional effects destabilizing  the jet \cite{ref:Planchette2018}. Second, for the points corresponding to $L_j/D_j>2$, the diameter ratios are slightly different with $D_j/D_d \approx 1.30$ here instead of $D_j/D_d \approx 1.19$ in \cite{ref:Baumgartner2020}. Yet, larger $D_j/D_d$ favours \textit{drops in jet}, limiting the jet fragmentation \cite{ref:Planchette2018}.  Experimental uncertainties can also play a role.  While the optical magnification used here corresponds to  not more than 6 $\mu m$ per pixel it was of approximately 10 $\mu m$ per pixel in \cite{ref:Baumgartner2020}, leading to uncertainties of up to $15\%$ for $We_d$ and $10\%$ for $L_j/D_j$. Other errors, such as the misalignment of the drop and jet trajectories in the same plane can of course not be totally excluded. Finally, it is also important to mention that the discrepancies concern a rather limited number of points (4 to 6).}

\begin{figure}
\centering   
\includegraphics[width=\textwidth]{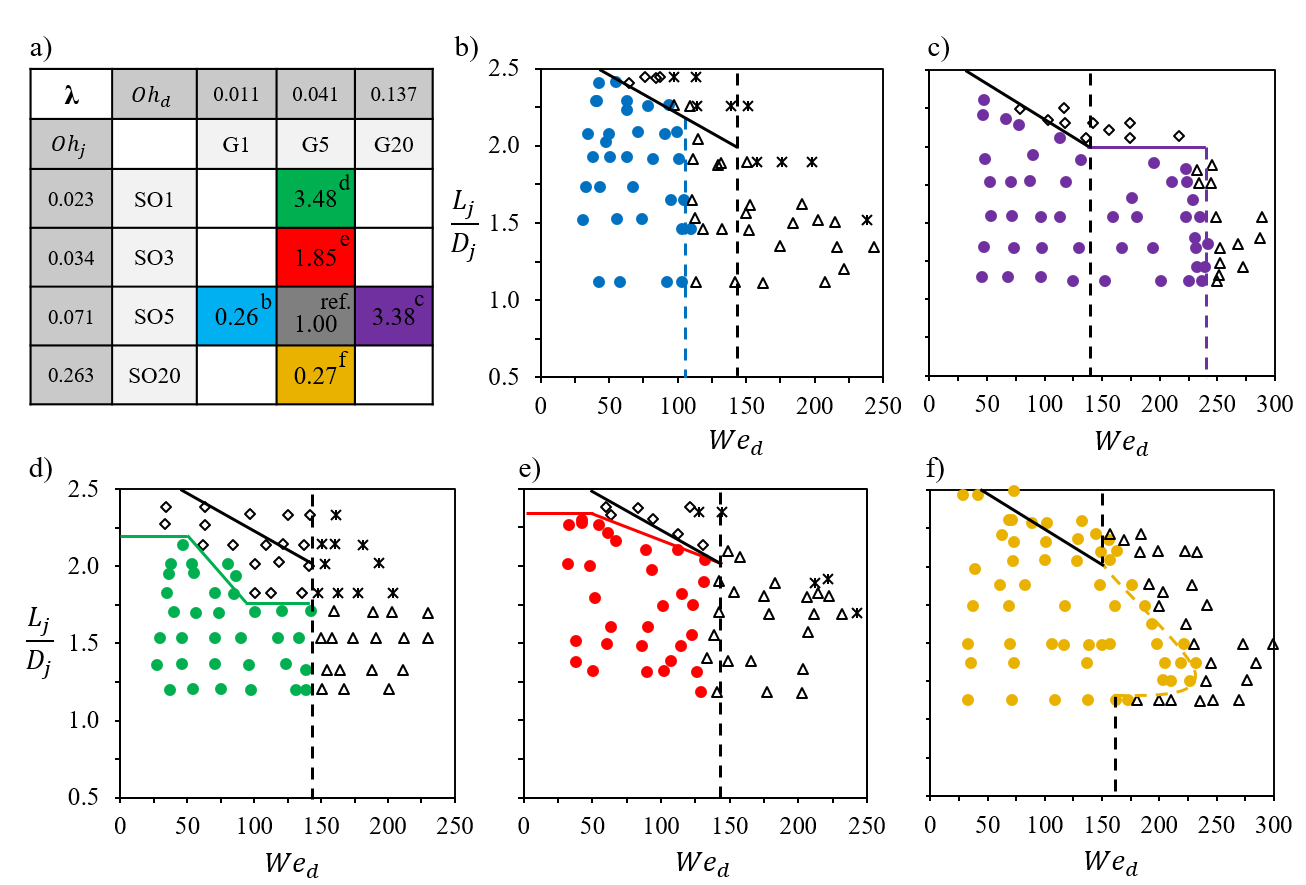}
\caption{(a)  {Studied liquid combinations with the corresponding drop / jet Ohnesorge numbers $Oh$  and associated viscosity ratio $\lambda=\mu_d/\mu_j$.} (b)-(f) The collision outcomes of the sets of experiments with changing drop or jet liquid in the form of regime maps with $We_d$ and $L_j/D_j$ as the scaling parameters - (b) drop liquid: G1, jet liquid: SO5; (c) drop liquid: G20, jet liquid: SO5; (d) drop liquid: G5, jet liquid: SO1; (e) drop liquid: G5, jet liquid: SO3; (f) drop liquid: G5, jet liquid: SO20. The coloured (solid or dashed) lines represent the boundaries for the respective liquid pair, while black lines indicate the boundaries for the reference case.}
\label{fig:regimemaps}
\end{figure}

The second scaling parameter is the drop Weber number $We_d=\rho_d D_d U^2 / \sigma_d$. It relates the kinetic energy of the impacting droplet to its surface energy. In agreement with other studies of various types of drop impact \cite{ref:Planchette2012, ref:Baumgartner2020, ref:Eggers2010, ref:Wildeman2016, ref:Lhuissier2013} it was found to quantify very well the drop deformation. It was further used to predict the inertial fragmentation of the drops, which distinguishes the \textit{drops-in-jet} and the \textit{fragmented drops-in-jet}. The transition between the two regimes is marked with a vertical dashed line (figures \ref{fig:referencecase} and \ref{fig:regimemaps})and indicates the maximum possible value of $We_d$ producing the stable \textit{drops-in-jet} structure. Further increase of drop inertia first leads to drop fragmentation and then to jet fragmentation, whatever the value of $L_j/D_j$. The critical drop Weber number of the reference case is $145\pm5$, and the transition from \textit{fragmented drops-in-jet} to \textit{mixed fragmentation} occurs at  $We_d\approx 180\pm5$. Note that this study especially focuses on the stable \textit{drops-in-jet} structure, since it is of big practical interest (coloured filled circles).  

Varying the drop viscosity while keeping the jet liquid unchanged (SO5), we obtain the results presented in figures \ref{fig:regimemaps}(b) and (c) corresponding to G1/SO5 and G20/SO5, respectively. Similarly, we then vary the jet liquid only and get the maps presented in figure \ref{fig:regimemaps}(d), (e) and (f) for SO1/G5, SO3/G5 and SO20/G5, respectively. The corresponding viscosity ratios $\lambda$ of the six liquid combinations are shown in figure \ref{fig:regimemaps}(a).

A first look at these regime maps (figures \ref{fig:regimemaps}(b)-(f)) reveals that the limits are not universal. Both the inertial and capillary fragmentation are shifted compared to the reference case. For better visualization of the shifts, the capillary (horizontal solid line) and inertial (vertical dashed line) fragmentation limits of the reference case G5/SO5 are indicated in black in the regime maps. Inevitably, the question arises: How does the viscosity of each liquid influence each fragmentation process?  The answer to this question is developed in the coming sections of this manuscript, based on a broad data set from more than 500 collision experiments. 
\section{Discussion}
\label{sec:discussion}
\subsection{Capillary fragmentation limit}
\label{sec:caplimit}
The capillary fragmentation limit corresponds to the transition between the continuous \textit{drops-in-jet} structure and the regular stream of \textit{encapsulated drops}. It is attributed to the fragmentation of the jet only and is therefore primarily considered within the range of stable drops, below their first inertial fragmentation. Observing figures \ref{fig:regimemaps}(b)-(e) demonstrates the relevance of $L_j/D_j$, which enables to clearly distinguish continuous and fragmented jet for all these liquid combinations. While the transition does not show significant deviations for varying the drop viscosity, it is considerably modified by the jet viscosity. Indeed, decreasing the jet viscosity (figure \ref{fig:regimemaps}(d)) destabilizes it, and capsules are formed for lower values of $L_j/D_j$. On the contrary, an increase of the jet viscosity leads to its stabilization and can even suppress the formation of capsules for values of $L_j/D_j$ up to 2.5 (figure \ref{fig:regimemaps}(f)). Above this value the jet may still fragment, but this breakup cannot be probed with our experimental set-up. 

To interpret these results, it may be useful to note that even for a given liquid pair the transition between the two regimes does not correspond to a unique value of $L_j/D_j$, as evidenced by the tilted solid lines. Furthermore, the critical range of $L_j/D_j$ is clearly found below the expected value of $\pi$ \cite{ref:Rayleigh1892} for all liquid combinations, except G5/SO20. In addition, a careful observation of the collisions suggests that the jet always fragments between two consecutive drops but not at the impacting points. This indicates that not the whole jet, but only its portions found between two drops are subjected to this capillary instability. As these jet portions are significantly distorted by the impacts, which stretched them in the direction of the drop momentum, it would explain why the critical values of $L_j/D_j$ are always below $\pi$, since this value does not account for the jet distortion. Further, the higher the drop momentum and thus $We_d$, the larger the stretching of the jet portions. As a result, the transition between continuous and fragmented jet occurs for critical values of $L_j/D_j$ slightly decreasing with increasing $We_d$. Keeping this in mind, we expect a secondary and limited effect of the drop viscosity on this transition. Indeed, the drop viscosity mainly affects the deformation of the drops and their immediate vicinity, but leaves the interspersed jet portions largely unaffected. This is in good agreement with the experimental results, which do not evidence significant modifications of the transition while varying the drop viscosity. 

In contrast, the jet viscosity is expected to play a major role. As the drops impact on the jet, they transfer a part of their momentum to the impacted parts of the jet. This transfer is of course not only a function of the drop velocity and mass (unchanged here, but studied in \cite{ref:Planchette2018}), but also of the jet viscosity. The latter causes losses, which limit the stretching of the jet. The greater the jet viscosity, the smaller the expected distortion of the interspersed jet portions and thus, the closer to $\pi$ the critical value of $L_j/D_j$. This is also in good agreement with the observations which are limited to $L_j/D_j\leq2.5$. To go further, it may be interesting to modify our experimental method in order to probe values of $L_j/D_j$ up to $\pi$, especially when viscous jets are used as, for example, G5/SO20 (see figure \ref{fig:regimemaps}(f)).

We are aware of the simplicity of the interpretation, and other second-order effects cannot be excluded.  Indeed, the considered situation is by far more complex than an ideal infinite liquid cylinder at rest. Momentum transfer, inertial flows and changes of the kinetics certainly contribute to the variations of the jet fragmentation. Yet, at first order attributing the fragmentation of the jet to the capillary instability of the portions found between two consecutive drops explains the decreasing values of $L_j/D_j$ with increasing $We_d$, the limited effect of the drop viscosity, and the major role of the jet viscosity.  

\subsection{Inertial fragmentation limit}
\label{sec:inertlimit}
Let us focus on the inertial fragmentation of the drops separating \textit{drops-in-jet} and \textit{fragmented drops-in-jet}. This inertial driven fragmentation mechanism is, in contrast to the capillary one, attributed to drop fragmentation only and may constitute a major limitation for applicability. The transition is marked with vertical dashed lines in all regime maps of figure \ref{fig:regimemaps}. A closer look at these lines immediately reveals that, by changing the drop liquid, the fragmentation limit is shifted. For low drop viscosity (G1/SO5; $\lambda=0.26$) the drop already breaks at $We_d=105\pm5$. In contrast, an increase of drop viscosity (G20/SO5; $\lambda=3.38$) raises the critical drop Weber number up to $204\pm5$. This result is indeed expected and in agreement with other studies which have revealed the effect of viscosity on the fragmentation of drops subjected to different types of impacts and collisions \cite{ref:Geppert2017, ref:Planchette2012, ref:Wildeman2016, ref:Eggers2010, ref:Walzel1980, ref:Shlegel2020}. Yet, in the given configuration where the drops impact on a jet, the understanding and quantification of the effect is missing. Regarding the effects of the jet viscosity, we observe only limited modifications of the transition when replacing SO5 by SO1 or SO3. Using SO20 leads to drop stabilization up to $We_d \approx 230$ for a certain range of $L_j/D_j$, approximately $1.25 \leq L_j/D_j \leq 1.75$. To understand these observations, we first focus on the evolution of the drop-jet system. Qualitative observations on the recorded collision pictures (orthogonal view) suggest that the overall evolution of the liquid system is similar to the one reported in \cite{ref:Baumgartner2020} and can be described as follows. First the drop forms a lamella, which reaches its maximum extension at $t_{d}^{max}$. This point is quickly followed by the maximum surface extension of the jet at $t_{j}^{max}$ which deforms in parallel. After that, the drop and jet recoil causing the drop to take an elongated shape. The latter fragments at $t_{frag}$ when its aspect ratio exceeds a critical value. The result is a continuous jet with a stream of encapsulated fragmented drops, each of them constituted of one main drop and two smaller ones originating from the extremities of the transiently elongated drop. We therefore develop an analysis starting from the effect of viscosity on the aspect ratio of the elongated drop, which we then try to explain by looking at the deformation of the drop.  

\subsubsection{Drop fragmentation}
\label{sec:drop_frag}
\begin{figure}
\centering 
\includegraphics[width=\textwidth]{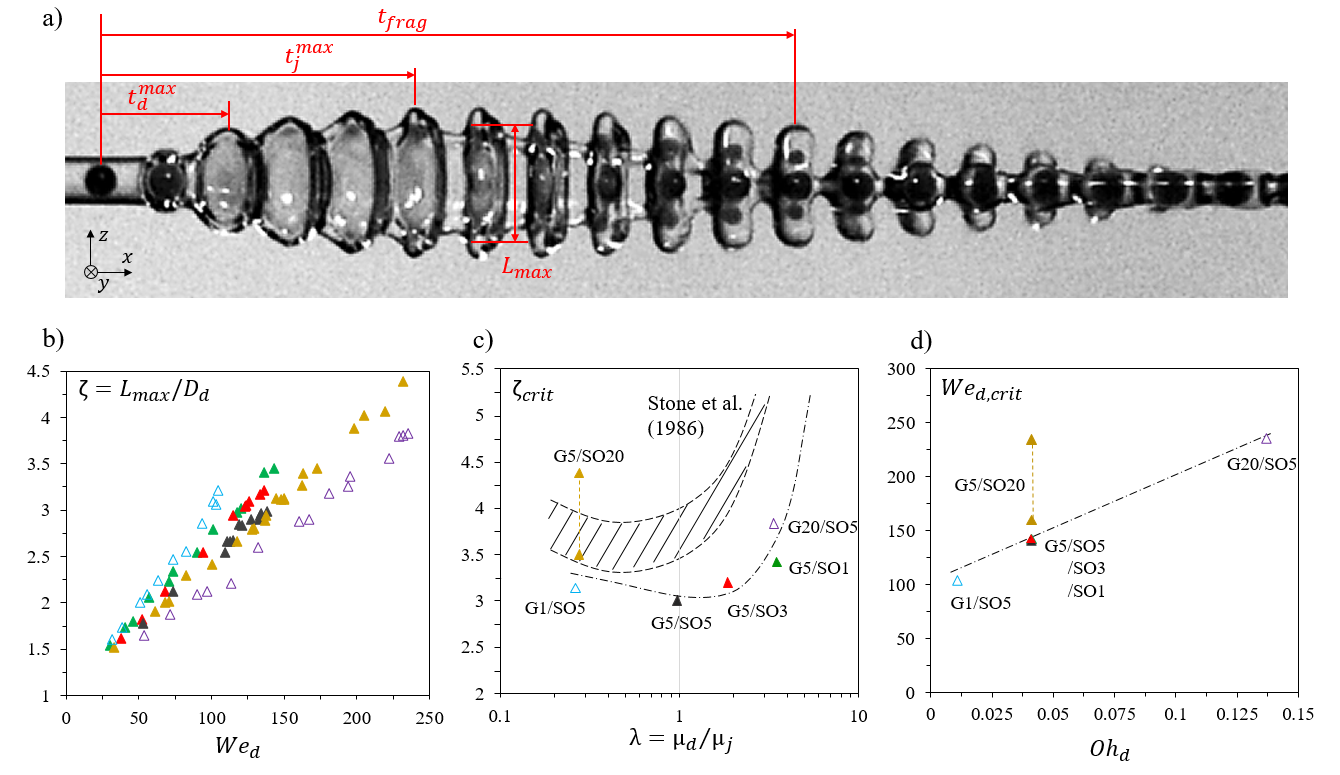}
\caption{(a) Recorded collision picture in orthogonal view with the three major instants at drop-jet-collisions and the maximum extension of the elongated drop  $L_{max}$.  {See supplementary materials for an illustrative movie \cite{ref:supplmat}.} (b) The aspect ratio $\zeta=L_{max}/D_d$ as a function of $We_d$ for all liquid pairs. (c) The critical aspect ratio $\zeta_{crit}$ at which fragmentation occurs against the viscosity ratio $\lambda$. {The dash-dotted line is a  fit function given by $\zeta_{crit}=0.007 \lambda^4 - 0.078 \lambda^3 + 0.422 \lambda^2 - 0.759\lambda+ 3.447$} (d) The critical Weber number $We_{d,crit}$ at which the drop fragments as a function of  { the drop Ohnesorge number $Oh_d$. The dash-dotted line is a guide for the eyes. (b-d) All the investigated configurations are accounted for corresponding to $1.1<L_j/D_j<2.3$.}}
\label{fig:fragmentation}
\end{figure}

Let us recall the evolution of the drop in more details. As for many other impacts, the drop deforms from an initial spherical shape to a bent lamella. After this bent lamella has reached its maximum surface extension, capillary forces drive the drop to recoil, as a liquid spring does \cite{ref:Planchette2012}. In case of drop-jet collisions, however, the drop does not isotropically contract inside the jet as for binary drop collisions or drop impact onto solids. Instead, it recoils in the direction of the jet axis (x-axis), while slightly further extending perpendicular to the jet, in the direction of the z-axis forming an elongated drop (like a cigar). For illustration, figure \ref{fig:fragmentation}(a) shows a collision picture recorded in orthogonal view with the instant of maximal drop extension $t_{d}^{max}$, the instant of maximal jet extension $t_{j}^{max}$ and the instant of drop fragmentation $t_{frag}$ as well as the maximum extension of the elongated drop $L_{max}$. Further, measurements confirm that the instant of drop fragmentation is reached after $t_{d}^{max}\approx 0.33t_{d,osc}$ at $t_{frag}\geq0.75t_{d,osc}$ for all liquid combinations and in agreement with the results of \cite{ref:Baumgartner2020}.
Here, $t_{d,osc}=(\rho_dD_d^3/\sigma_d)^{1/2}$ stands for the capillary time of the droplet. The maximum extension $L_{max}$ normalized by the initial drop diameter $D_d$ gives the aspect ratio $\zeta=L_{max}/D_d$, which is shown in figure \ref{fig:fragmentation}(b) as a function of $We_d$ for all six liquid combinations. If the aspect ratio exceeds a critical value $\zeta_{crit}$, end-pinching occurs and the drop fragments. The extremities of the dumbbell-shaped drop detach from the central part, which contracts back to a sphere.  {The critical values $\zeta_{crit}$  are plotted against the viscosity ratio $\lambda$ in figure \ref{fig:fragmentation}(c). To provide an overview of the transition, the corresponding critical Weber numbers $We_{d, crit}$ are reported as a function of the drop Ohnesorge numbers $Oh_d$ in figure  \ref{fig:fragmentation}(d). } Black full triangles represent the reference case G5/SO5. Fully coloured symbols represent liquid combinations with unchanged drop liquid (G5) and various jet liquids, while empty symbols stand for unchanged jet liquid (SO5) and different drop liquids. First, note that the values of $We_{d,crit}$ reported in figure \ref{fig:fragmentation}(d)  {as a function of $Oh_d$} confirm the transition shifts observed in the maps of figure \ref{fig:regimemaps}. While the variation of the drop viscosity (empty symbols) considerably affects $We_{d,crit}$, reducing the jet viscosity from 5 mPa$\cdot$s to 3 and 1 does not trigger significant differences  {(see full symbols close to $We_{d,crit}=140$ for $Oh_d \approx 0.04$)}. The increase of the jet viscosity up to 20 mPa$\cdot$s broadens the transition, which does not occur for a unique value of $We_d$ anymore. Depending on $L_j/D_j$, this fragmentation takes place for $155\leq We_d \leq 230$, as illustrated by the dashed line connecting the  {upper full symbols at $Oh_d \approx 0.04$. This phenomenon is attributed to a change of  fragmentation mechanism and is discussed at the paper end}. These results also  {confirm} the fact that the drop and the jet viscosity do not play the same role  { and validate \textit{a posteriori} the empirical use of $Oh_d$. Yet,} keeping in mind that the drop fragmentation was attributed to a critical aspect ratio of the encapsulated drop during its recoil, and that, as reported by Stone et al.  {for quiescently elongated drops} \cite{ref:Stone1986, ref:Stone1989}, Stone \cite{ref:Stone1994} and Grace \cite{ref:Grace1982}, the latter is a function of the viscosity ratio, it is legitimate to ask  {how these findings could apply for drop-jet collisions and if the inertial transition empirically described by   $We_{d,crit}$ with $Oh_d$ could not be better predicted}. 
\FloatBarrier

To answer this question, it is helpful to look at figure \ref{fig:fragmentation}(c), where the variations of $\zeta_{crit}$ with $\lambda$ are reported together with the  previously reported results of Stone et al. \cite{ref:Stone1986} (hatched area). Here again, the empty symbols correspond to SO5 combined with different drop liquids, while the full ones show unchanged drop liquid combined with the jet of various viscosities. This figure demonstrates that the drop fragmentation does not occur always for the same critical value of $\zeta$ as the one reported in \cite{ref:Baumgartner2020}. It also shows that, in contrast to the variations of $We_{d,crit}$ with $\lambda$, both the drop and the jet viscosity affect this critical value. As reported by Stone et al., the minimum $\zeta_{crit}$ is reached for $\lambda=1$. In other words, matching the drop and jet viscosity leads to the less stable "cigar". The critical values found for drop-jet collisions are systematically lower than those evidenced for elongated ligaments in quiescent fluids. Yet, the difference remains moderate and may easily be explained by the flow originating from the contracting jet and by the significant deviations of the elongated drop from a perfect cylinder. The values obtained for a given $\lambda$  {seem to be} different, depending if the encapsulated drop or the surrounding jet viscosity is varied. In Stone et al. \cite{ref:Stone1986} only the viscosity of the elongated cylinder was changed, which makes the comparison impossible.  {Nevertheless, they also observed a rather large range of  values (see hatched area) so that the dispersion may have another origin.} The increase of $\mu_j$ to 20 mPa$\cdot$s causes  {a further broadening of the data, the value of $\zeta_{crit}$ becoming a function of $L_j/D_j$. This may be caused by the emergence of a different fragmentation mechanism which is discussed at the end of the paper. Thus, despite the mentioned data dispersion, we propose to approach the variations of $\zeta_{crit}$ with $\lambda$ by the following fitting function: $\zeta_{crit}=0.007 \lambda^4 - 0.078 \lambda^3 + 0.422 \lambda^2 - 0.759\lambda+ 3.447$, see the dash-dotted line in figure \ref{fig:fragmentation}(c).} 

To go further and relate the critical values of $\zeta$ to the ones  {for example} of $We_d$, we  {propose  to  investigate the evolution of $\zeta$ with $We_d$.} The curves corresponding to each liquid combination are shown in figure \ref{fig:fragmentation}(b). The full symbols representing different jet liquids remain close to each other, demonstrating that the jet viscosity is not the key parameter fixing the shape taken by the recoiling drop. Its aspect ratio can be well described by $\zeta=aWe_d+b$, where $a=0.0155\pm0.0013$ and $b=1.0193\pm0.0375$ are deduced from a fitting procedure applied to each curve (see graphs in supplementary materials \cite{ref:supplmat}). In contrast, the drop viscosity significantly modifies the evolution of $\zeta$ with $We_d$, so that, the smaller the viscosity, the greater the slope ($a_{G1}=0.0211;\,a_{G20}=0.0121$; graphs shown in supplementary materials \cite{ref:supplmat}). These observations call for a deeper analysis of the drop deformation, which is proposed in the next section.         

\subsubsection{Drop deformation}
\label{sec:dropextension}
The bent lamella forming after the drop impact on the jet can be modelled as a flat disk. The diameter of the disk $D_{lam}$ grows until the maximum extension of the drop is reached, corresponding to $D_{lam}^{max}$ and to a maximal surface of the drop $\Sigma_{d}^{max}$. By analogy with a liquid spring \cite{ref:Planchette2017}, this initial phase is named the compression phase. After this instant, the drop contracts in order to minimize its interfacial area. To evaluate the drop deformation, we extract from the images in front and orthogonal views the temporal evolution of the lamella surface  $\Sigma_{d,lam}$ for each collision. Detailed information about this procedure is reported in sections 2.3 and 4.1 of \cite{ref:Baumgartner2020}. In short, combining the volume conservation of the drop before and after impact with $D_{lam}$ taken from the pictures leads to the expression \cite{ref:Planchette2017, ref:Baumgartner2020}
\begin{equation}
\Sigma_{d,lam}=\frac{\pi D_{lam}^2}{2} \biggl(1+\frac{8}{3}\frac{D_d^3}{ D_{lam}^3}\biggr)
\label{eq:maxdropsurface}
\end{equation}
\FloatBarrier
\begin{figure}
\centering 
\includegraphics[width=\textwidth]{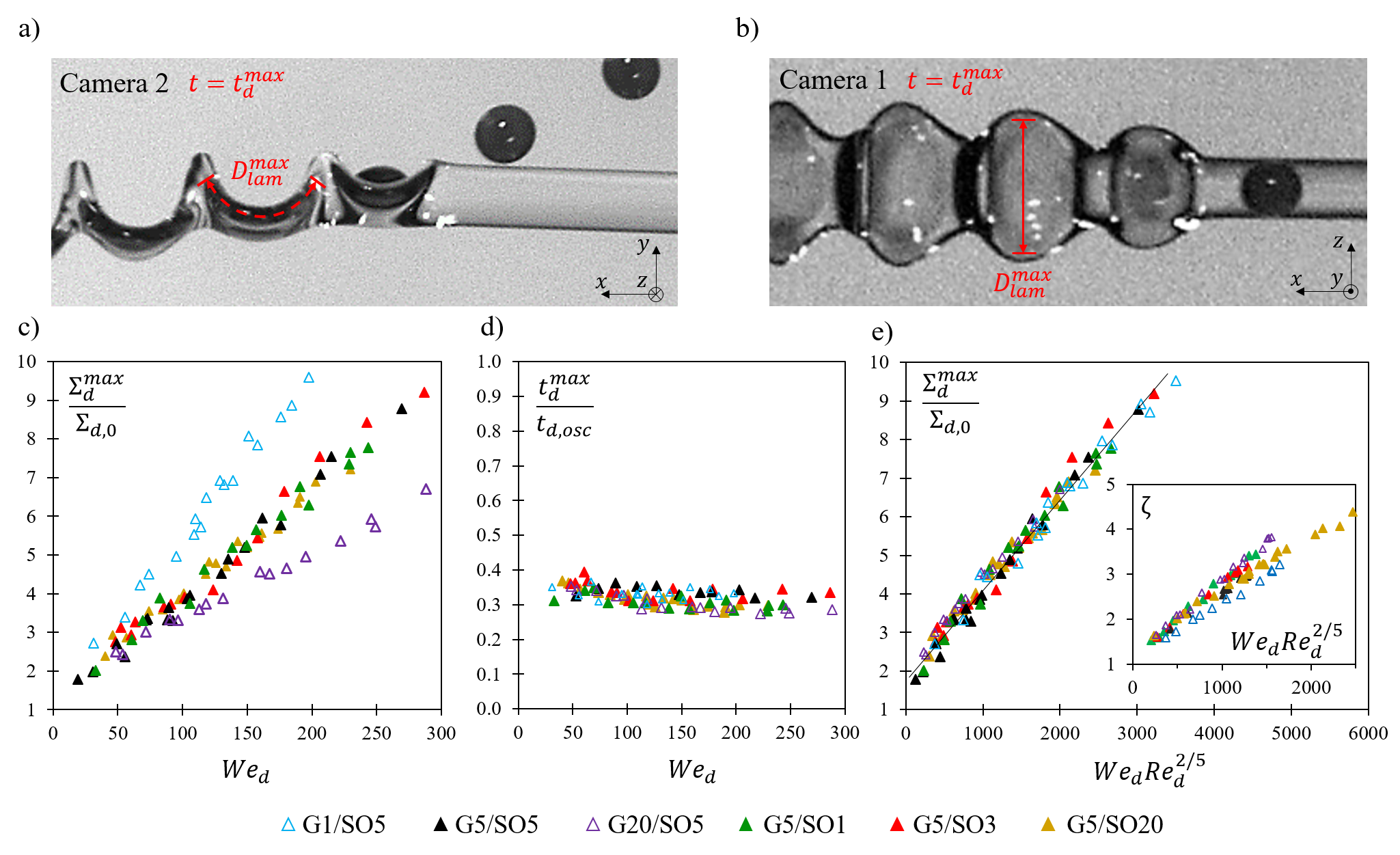}
\caption{Collision pictures in (a) front view and (b) orthogonal view with $D_{lam}^{max}$, the maximum diameter of the bent lamella, and $t_{d}^{max}$, the instant of maximal extension. (c) The maximum surface of the deformed drop  $\Sigma_{d}^{max}$ normalized by $\Sigma_{d,0}$ as a function of $We_d$ for all six liquid combinations. (d) Instant of maximal extension $t_{d}^{max}$ normalized by $t_{d,osc}$. (e) Normalized maximum drop surface $\Sigma_{d}^{max}/\Sigma_{d,0}$ as a function of $We_dRe_d^{2/5}$. Inset: The aspect ratio $\zeta$ as a function of $We_dRe_d^{2/5}$.}
\label{fig:dropextension}
\end{figure}

By fitting the temporal evolution of $\Sigma_{d,lam}$ with a parabola, the maximum drop surface  $\Sigma_{d}^{max}$ and the instant of maximal extension $t_{d}^{max}$ are obtained. Pictures recorded at  $t_{d}^{max}$ are represented in figure \ref{fig:dropextension}(a) in front view (camera 2) and (b) in orthogonal view (camera 1) and schematically show $D_{lam}^{max}$, which is otherwise extracted from the parabolic fit. $\Sigma_{d}^{max}$ is normalized by the initial drop surface  $\Sigma_{d,0}=D_d^2 \pi$ and plotted against $We_d$ in figure \ref{fig:dropextension}(c). Similarly, $t_{d}^{max}$ normalized by the capillary time of the droplet $t_{d,osc}=(\rho_d D_d^3/\sigma_d)^{1/2}$ is shown as a function of $We_d$ in figure \ref{fig:dropextension}(d). 

The time of maximal extension is independent of both the drop and jet viscosity, as well as the kinetic energy of the drop. It is fixed by a purely capillary scaling, as already found for miscible and immiscible drop-jet collisions \cite{ref:Baumgartner2020}, miscible and immiscible drop-drop collisions \cite{ref:Planchette2017}, and for the contact time of a bouncing drop \cite{ref:Richard2002}. The maximum drop surface scales very well with $We_d$, which is in good agreement with observations of other drop impact processes, e.g. binary drop collisions \cite{ref:Roisman2012}, drop impacts on a solid wall \cite{ref:Huang2018} or drop-jet collisions with partially wetting immiscible liquids \cite{ref:Baumgartner2020}. Nevertheless, a difference in slope can be seen for changing drop liquid. Confirming the observations of $\zeta=f(We_d)$, droplets with low viscosity spread significantly more than those of high viscosity. Similar results were reported by Laan et al. \cite{ref:Laan2014} for drop impacts on a hard surface. Here, droplets made of the same liquid (G5) extend equally, independently of the surrounding jet liquid and its viscosity. Qualitatively, the evolution of the drop extension $\zeta$ appears very similar to the one of the maximum surface of the encapsulated droplet and is therefore worth being understood.  

A simple picture of drop extension consists in considering that, at maximum extension, the initial kinetic and surface energy of the drop is converted into surface energy, either with or without losses, corresponding to either a capillary-inertial regime or a viscous one \cite{ref:Roisman2009a, ref:Roisman2009b, ref:Wildeman2016, ref:Eggers2010}. The former comes to play when viscosity effects can be neglected. The balance between inertia and surface tension then defines the drop spreading, which scales as  $\Sigma_{d}^{max}/ \Sigma_{d,0} \propto We_d$ \cite{ref:Eggers2010}. The second regime must be considered when viscous dissipation dominates surface tension. In this case, most of the initial drop kinetic energy is converted into viscous losses, and the maximum drop spreading can be obtained with  $\Sigma_{d}^{max}/ \Sigma_{d,0} \propto Re_d^{2/5}$ \cite{ref:Eggers2010}, where $Re_d=D_dU/\nu_d$ is the drop Reynolds number. Here, neither $We_d$ (see figure \ref{fig:dropextension}(c)) nor $Re_d^{2/5}$  (data not shown) can be used to define a satisfactory scaling which suggests that an interplay between inertial, surface and viscous forces has to be considered. Different studies, for instance of binary drop collisions \cite{ref:Roisman2004}, drop impact onto solid walls \cite{ref:Roisman2009a, ref:Roisman2009b, ref:Wildeman2016, ref:Eggers2010}, drop impact onto a spherical target \cite{ref:Bakshi2007} and also drop-jet collisions with  immiscible partially wetting liquids \cite{ref:Baumgartner2020}, show that the energy balance approach is not able to describe the flow inside the deforming droplet and can therefore not be used to predict the maximum drop extension. Instead let us consider a momentum balance as in \cite{ref:Ashgriz1990}. Following this approach, and given the absence of external forces, the drop stops deforming and reaches its maximum extension when the initial kinetic pressure $p_{kin}$ acting on $\Sigma_{d,0}$ is sustained by the internal pressure p acting on the deformed surface $\Sigma_{d}^{max}$. Since the droplets do not seem influenced by the jet liquid within the compression phase (figure \ref{fig:dropextension}(c)), we derive these forces based on the analogy with the well studied process of drop impact on a solid substrate \cite{ref:Roisman2009a, ref:Roisman2009b, ref:Wildeman2016, ref:Eggers2010}. The jet, whose mass and momentum are greater than those of the drops, play the role of the immobile solid substrate which the drop impacts with the (relative) velocity U. Using cylinder coordinates \{r, $\varphi$, y\}, the internal pressure of the liquid sheet at maximum extension is given by \cite{ref:Roisman2009a, ref:Roisman2009b}
\begin{equation}
\sigma_{rr}+\sigma_{\varphi \varphi}+\sigma_{yy}=-3p
\label{eq:internalpressure}
\end{equation}
with, by accounting for axisymmetric drop extension, the stress tensor components reading:
\begin{equation}
\sigma_{rr}=-p+2\mu_d \frac{\partial v_r}{\partial r}, \qquad \sigma_{\varphi \varphi}=-p+2\mu_d \frac{v_r}{r}, \qquad \sigma_{yy}=-p_{\sigma}.
\label{eq:stresscomponents}
\end{equation}
Here $v_r(r,t)$ is the radial internal velocity component. $\sigma_{yy}$ corresponds to the capillary pressure jump given by the Young-Laplace equation $p_{\sigma}\approx -\sigma_d [\frac{1}{r} \frac{\partial h}{\partial r} + \frac{\partial^2 h}{\partial r^2}]$ where $h(r,t)$ is the overall thickness of the spreading lamella. Thanks to the very thin and flat shape of the lamella yielding $\partial h / \partial r \ll 1$, it can be approximated by $p_{\sigma}\approx -\sigma_d [\frac{\partial^2 h}{\partial r^2}]$. Including the conditions $v_r\approx0$ and $\partial{v_r}/\partial{r}\approx0$ at $t_{d}^{max}$ (no internal flow at the instant of maximum extension) leads to the fact that the capillary pressure is the primarily existing pressure in the liquid sheet at maximum extension, i.e. $p=p_{\sigma}$ at $t=t_{d}^{max}$. Thus, we deduce
\begin{equation}
\frac{\Sigma_{d}^{max}}{\Sigma_{d,0}} = We_d \biggl[\frac{\partial^2 h^*}{\partial {r^*}^2}\biggr]^{-1}
\label{eq:drop_scaling_01}
\end{equation}
with $h^*=h/D_d$ the dimensionless lamella thickness. Note, that here and in the rest of the manuscript the dimensionless form of each parameter is indicated with $^*$, using $U$ as the velocity scale and $D_d/U$ as the time scale. Making further use of the analogy with a drop impacting on a solid, and considering for example the findings of Rioboo et al. \cite{ref:Rioboo2002}, we expect the shape of the lamella to be significantly influenced by the drop viscosity, and thus $h^*$ to be a function of $Re_d$, the drop Reynolds number. This viscous effect can be explained by the development of a viscous boundary layer near the wall and the formation of a rim at the edge of the lamella. The viscous boundary layer thickness $\delta^*$ can first be estimated by $\delta^* \approx \sqrt{t^*/Re_d}$ \cite{ref:Roisman2009b, ref:Wildeman2016, ref:Eggers2010}. As long as $\delta^* < h^*$, the dimensionless thickness of the spreading film can be well described by the remote asymptotic solution for inertia dominated inviscid flow, providing  $h^*=h^*_{inv}(t)+h^*_{\nu}(t)$. Here, $h^*_{inv}$ represents the inviscid part and $h^*_{\nu}$ stands for the viscous thickness, which can be neglected for early stages and high drop Reynolds numbers \cite{ref:Roisman2009b}. 

When the thickness of the boundary layer reaches the lamella thickness, a different solution must be considered. In this case, and as initially derived for drop impact onto spherical substrates \cite{ref:Bakshi2007}, the film thickness is the solution of
\begin{equation}
\ddot{h}^*- \frac{9}{5} \frac{\dot{h}^{*2}}{h^*}+ \frac{3}{Re_d} \frac{\dot{h}^*}{h^{*2}}=0
\label{eq:lamellathickness02}
\end{equation}
and reaches some asymptotic value, which is defined as the residual film thickness given by
\begin{equation}
h^*_{res}=\frac{h^{*9/14}_0}{\biggl( \frac{1}{h^*_0}+\frac{14Re_dV^*_0}{15} \biggr)^{5/14}}.
\label{eq:resthickness}
\end{equation}
Here, $h^*_0$ represents the thickness of the lamella at the beginning of this phase, when the boundary layer reaches the free drop surface and $V^*_0=- d h^* / d t^*$ stands for the rate of film thinning at this instant.

Coming back to drop impact onto a liquid jet, and making further use of the analogy with solid substrates, we estimate the dimensionless time at which the boundary layer reaches the free drop surface by $t^*_{BL} \approx  0.59 Re_d^{1/5}$ \cite{ref:Roisman2009b}. In this study, all drop-jet collisions proceed by the full development of the boundary layer at $t^*_{BL}$ first, followed by the maximum extension of the drop at ${t_{d}^{max}}^*$. In other words, the condition ${t_{d}^{max}}^*>t^*_{BL}$ is always fulfilled. This reveals that the lamella thickness can be well estimated by the residual film thickness of equation (\ref{eq:resthickness}). Relating thereto, and by analogy with drop impacts on solids, the conditions for the initial thickness of the lamella $h^*_0$ and the initial rate of film thinning  $V^*_0$ at $t^*=t^*_{BL}$ can be approximated by $h^*_0 \approx 1.46 Re_d^{-2/5}$ and $V^*_0 \approx 3.31 Re_d^{-3/5}$ \cite{ref:Roisman2009b}. This leads to a residual film thickness, which only scales with the drop Reynolds number as 
\begin{equation}
h^*_{res} \approx 0.79Re_d^{-2/5}
\label{eq:residualthickness}
\end{equation}
Finally, replacing the dominating curvature ${\partial^2 h^*}/{\partial {r^*}^2}$ in equation (\ref{eq:drop_scaling_01}) by $1/h^*_{res} \approx Re_d^{2/5}$, we obtain
\begin{equation}
\frac{\Sigma_{d}^{max}}{\Sigma_{d,0}} = \alpha_d We_d Re_d^{2/5}  + \beta_d
\label{eq:drop_scaling_02}
\end{equation}
This scaling is applied to all the collisions of this study, and the results are shown in figure \ref{fig:dropextension}(e). All experimental data are brought together in very good agreement with the proposed scaling. This confirms the interest of momentum balance for drop impacts and reveals that the drop extension is fixed by a capillary-inertial balance affected by the drop viscosity which fixes the lamella thickness. The solid line of figure \ref{fig:dropextension}(e) corresponds to equation (\ref{eq:drop_scaling_02}) with $\alpha_d = 0.0023$ and $\beta_d = 1.901$. The fact that $\beta_d \neq 1$ indicates a limited validity of this scaling for low inertia, which is expected since the boundary layer cannot fully develop. Similar deviations have been reported for binary and ternary immiscible drop collisions \cite{ref:Planchette2012, ref:Planchette2017}.

\subsubsection{ {Predicting the fragmentation threshold}}

As already mentioned,  the trends observed for drop extension (figure \ref{fig:dropextension}(c)) are quite similar to those of $\zeta$ (figure \ref{fig:fragmentation}(b)). Indeed, the jet viscosity seems to play a limited role, and the more viscous the drop, the smaller the deformation and the aspect ratio.

\begin{figure}
\centering 
\includegraphics[width=0.66\textwidth]{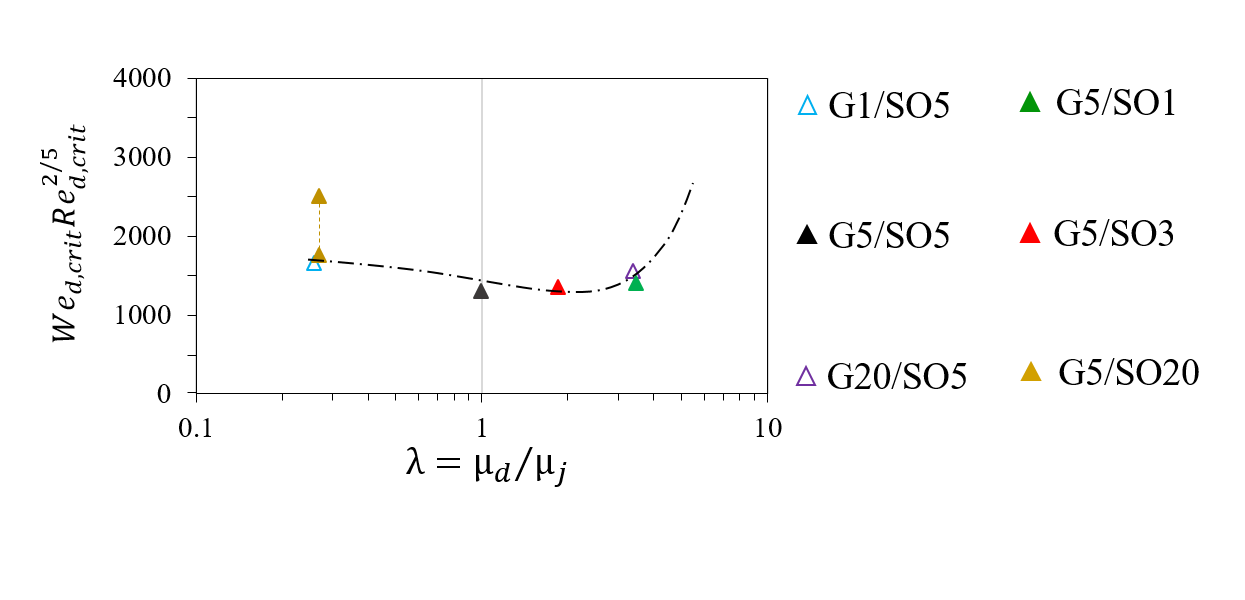}
\caption{ {Inertial fragmentation threshold in the form of  $We_{d,crit} Re_{d,crit}^{2/5}$ as a function of $\lambda$. Symbols: experimental data, dash dotted line: Eq. (\ref{eq:thresold}).}}
\label{fig:We_drop_Ohnesorge}
\end{figure}

 {We thus propose to make use of the drop extension findings to predict the evolution of $\zeta$. Employing the same scaling, we obtain a rather good agreement with the experimental data, see insert figure \ref{fig:dropextension}(e). 
Further combining this scaling with the function used to describe the evolution of $\zeta_{crit}$ with $\lambda$ provides a fragmentation threshold in the form of:}
\begin{equation}
\label{eq:thresold}
    We_{d,crit}{Re_{d,crit}}^{2/5} = 119 {\lambda}^2 - 500 {\lambda} + 1800 
\end{equation}
 {This prediction is plotted in figure \ref{fig:We_drop_Ohnesorge} (dash-dotted line) together with the experimental data (symbols). The agreement is reasonably good, but deviations can be observed which can easily be explained. First, the variations of $\zeta_{crit}$ with $\lambda$ show a rather important dispersion which reduces the relevance of the fitting function $\zeta_{crit}=f(\lambda)$ (figure \ref{fig:fragmentation}(c)) and thus the one of Eq. (\ref{eq:thresold}). Second, the scaling in $We_dRe_d^{2/5}$ which shows excellent agreement for the drop extension  $\Sigma_{d}^{max}/\Sigma_{d,0}$, has a limited validity when applied to $\zeta$, see figure \ref{fig:dropextension}(e). Indeed, the effects of drop viscosity appear to be slightly over-predicted, while those of the jet are not accounted for. In this context, the empirical usage of a simple parameter such as the drop Ohnesorge number $Oh_d$ may be of interest, see fig \ref{fig:fragmentation}(d).}
 {To conclude, }the drop viscosity mainly affects the inertial transition via the extension of the drop, which itself is influenced by the thickness of the boundary layer developing during the first instants of the lamella extension. The greater the drop extension, and the greater the aspect ratio of the encapsulated drop, which facilitates its pinch-off. Further we show that in agreement with the findings of Stone et al. \cite{ref:Stone1986}, matching the drop viscosity with the one of the jet leads to the least stable elongated drop. A more viscous drop results in higher values of $\zeta_{crit}$ and therefore enhances the stabilization obtained by the reduction of $\zeta$. Except for SO20, which is discussed separately, the jet viscosity does not really affect this transition, at least not in terms of the associated $We_{d,crit}$. Looking more carefully, we see that the jet viscosity modifies $\zeta$, and thus the geometry of the recoiling drop, but these effects are primarily compensated by the associated variations of $\zeta_{crit}$ with $\lambda$. This compensation may originate from the fact that the jet deformation is mainly driven by the drop movements during both the extension phase and the recoil. The viscous effects experienced in these two phases may therefore cancel each other, so that the drop extension alone determines the transition. Indeed, we measure greater extension of the jet surface for less viscous jet liquids (see supplementary materials \cite{ref:supplmat}) and can similarly expect greater jet recoil for less viscous jets. Note that modelling the jet extension goes beyond the scope of this paper and would not provide a better understanding of the compensating effects.
\subsubsection{ {Case G5/SO20}}
\begin{figure}
\centering 
\includegraphics[width=0.8\textwidth]{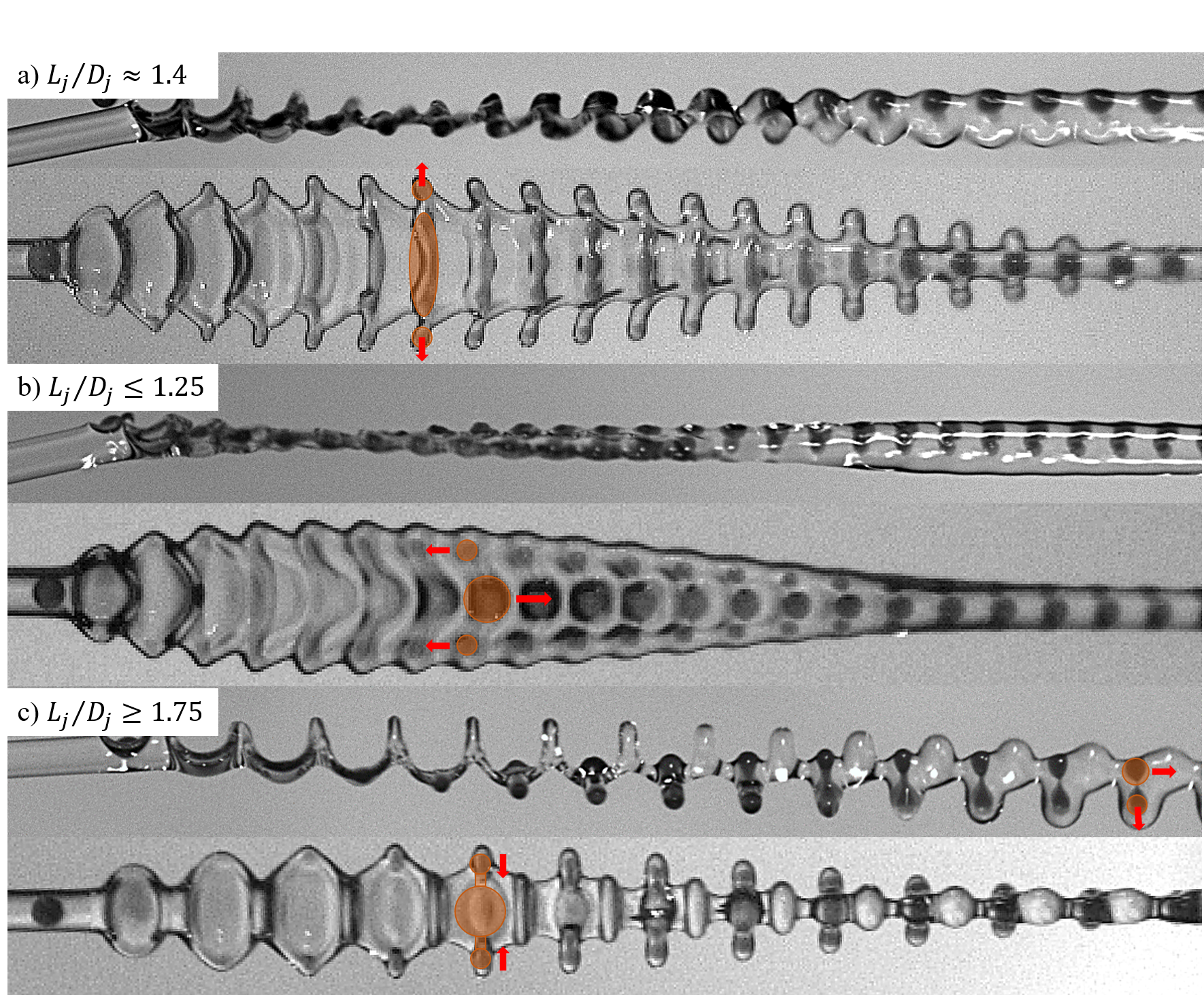}
\caption{Collision pictures illustrating the three droplet fragmentation processes observed for G5/SO20 in front (upper picture) and orthogonal (picture below) view: (a) $L_j/D_j \approx 1.4$ (b) $L_j/D_j \leq 1.3$ (c) $L_j/D_j \geq 1.75$.}
\label{fig:drop_frag_SO20}
\end{figure}
Let us now come back to the combination of G5/SO20. In contrast to the other combinations studied and to the results of the literature \cite{ref:Baumgartner2020}, the transition between \textit{drops-in-jet} and \textit{fragmented drops in jet} does not correspond to a unique value of $We_{d,crit}$. The minimum ($We_{d,crit} \approx 155$) being found for small and large $L_j/D_j$ ($L_j/D_j \leq 1.25$ and $L_j/D_j \geq 1.75$), the maximum ($We_{d,crit} \approx 230$) is around $L_j/D_j \approx 1.4$. We explain this phenomenon by the existence of three unequal fragmentation processes, which are illustrated in figure \ref{fig:drop_frag_SO20}. First, in the case of intermediate $L_j/D_j \approx 1.4$, we recognize the typical evolution taking place with other combinations. The encapsulated drop is elongated in the lamella plane, perpendicularly to the jet axis, and fragments at its extremities if exceeding a critical aspect ratio of $\zeta_{crit} \approx 4.3$, leaving a main drop and two satellites, indicated with orange circles in figure \ref{fig:drop_frag_SO20}(a). The red arrows symbolize the movements of the drop fragments. Second, for small $L_j/D_j \leq 1.25$, the drop extremities pinch off at smaller $We_{d,crit}$ and at smaller $\zeta_{crit}$. This can be explained by the interaction of two consecutive drops. The drop extends and recoils at first, similarly to the one at larger $L_j/D_j$. At a certain point in the recoiling phase, however, the extension of the following droplet forces the central part of the recoiling droplet to move further on in the direction of the jet axis, while the drop extremities keep their positions. A shearing effect is induced by this relative displacement, leading to premature pinch-off observed at smaller values of $We_{d,crit} \approx 170$ and  $\zeta_{crit} \approx 3.43$ (figure \ref{fig:drop_frag_SO20}(b)). Finally, in the case of larger $L_j/D_j \geq 1.75$, the shape of the liquid system does not correspond to a succession of elongated recoiling drop connected by narrower parts made of the jet liquid only (figure \ref{fig:drop_frag_SO20}(c)). The jet envelope remains rather constant in section but develops bulges in the direction of the relative velocity. The droplet extends and recoils again, similarly to all other collision experiments which may originate from a reduced momentum exchange between the drop and the jet. Yet, the bulges developing in the collision plane force  the elongated drop to stretch normally to its main axis. Shearing arises, which facilitates the pinch-off of the dumbbell-shaped drop. As a result, the fragmentation occurs at smaller $We_{d,crit} \approx 155$, while $\zeta_{crit} \approx 3.21$ is underestimated due to its bend. 
While the quantification of these different processes remains a challenge, our study shows that, using jets three times more viscous than the drop, it may be needed to account for $L_j/D_j$ in order to predict the inertial drop fragmentation.        
\section{Conclusions}
We have experimentally probed the effects of viscosity on the structures produced by drop-jet collisions. To do so, drops from glycerol solutions were combined with jets from various silicone oils, providing viscosity ratios $\lambda=\mu_{d}/\mu_{j}$ between 0.25 and 3.50. We observe that viscous jets (G5/SO20) could suppress their fragmentation in the range of studied parameters. More generally, and whatever the drop viscosity, the more viscous the jet, the more stable it is, enlarging the domain of \textit{drops-in-jet} at the expense of the capsule regime. We attribute the jet fragmentation to a pseudo Rayleigh-Plateau instability of the jet sections between two consecutive drops. Thus, the more viscous the jet, the less elongated these sections and the greater the drop spacing must be to lead to jet fragmentation. Further, we demonstrate that the jet viscosity does not affect the drop fragmentation, which is mainly influenced by the drop viscosity. Only in case of a jet three times more viscous than the drop effects could be noticed. The otherwise observed evolution of the recoiling system was replaced for small and large $L_j/D_j$ by different phenomena associated to shearing and shifting the fragmentation limit. We attribute this transition separating \textit{drops-in-jet} and \textit{fragmented drops-in-jet} to end-pinching of the recoiling encapsulated drop. As for a perfect cylinder, the fragmentation of the drop corresponds to a critical value of its aspect ratio. This aspect ratio results mainly from the extension of the drop, which can be very well described as function of $We_d Re_d^{2/5}$. We establish this scaling based on a momentum balance accounting for the viscosity via the thickness of the boundary layer that develops during the lamella extension. Thus, the more viscous the drop, the smaller the aspect ratio reached by the encapsulated drop. We have additionally shown that the stability of the drop is further enhanced by the evolution of the critical aspect ratio value with the viscosity ratio $\lambda$. These results are in agreement with those of Stone et. al \cite{ref:Stone1986} and may be used to favour the \textit{drops-in-jet} regime.


\begin{thebibliography}{99}

\bibitem{ref:Rosen2005}H. Rosen, and T. Abribat, The rise and rise of drug delivery, \textit{Nat. Rev. Drug Discov.}, 4, pp. 381–385 (2005).

\bibitem{ref:Kumari2010}A. Kumari, S.K. Yadav, and S.C. Yadav, Biodegradable polymeric nanoparticles based drug delivery systems, bioengineered and biomimetic drug delivery carriers, \textit{Colloids Surf. B}, 75, pp. 1-18 (2010).

\bibitem{ref:Kearney2013}C. Kearney, and D. Mooney, Macroscale delivery systems for molecular and cellular payloads, \textit{Nat. Mater}, 12, pp. 1004–1017 (2013).

\bibitem{ref:Yoo2011}J.-W. Yoo, D.J. Irvine, D., D.E. Discher, and S. Mitragotri, Bio-inspired, bioengineered and biomimetic drug delivery carriers, \textit{Nat. Rev. Drug Discov.}, 10, pp. 521–535 (2011).


\bibitem{ref:Bhatia2005}S.R. Bhatia, S.F. Khattak, and S.C. Roberts, Polyelectrolytes for cell encapsulation, \textit{Curr. Opin. Colloid Interface Sci}, 10, pp. 45-51 (2005).

\bibitem{ref:Nicodemus2008}G.D. Nicodemus, and S.J. Bryant, Cell encapsulation in biodegradable hydrogels for tissue engineering applications, \textit{Tissue Eng. Part B Rev.}, 14, pp. 149-165 (2008).

\bibitem{ref:Wang2014}S. Wang, and Z. Guo, Bio-inspired encapsulation and functionalization of living cells with artificial shells, \textit{Colloids Surf. B}, 113, pp. 483 - 500 (2014).

\bibitem{ref:Steele2014}J.A.M. Steele, J.-P. Hallé, D. Poncelet, and R.J. Neufeld, Therapeutic cell encapsulation techniques and applications in diabetes, \textit{Adv. Drug Deliv. Rev.}, 67-68, pp. 74 - 83 (2014).

\bibitem{ref:Yeo2004}Y. Yeo, A.U. Chen, O.A. Basaran, and K. Park, Solvent exchange method: a novel microencapsulation technique using dual microdispensers, \textit{Pharm. Res.}, 21, pp. 1419-1427 (2004).

\bibitem{ref:Brandenberger1998}H. Brandenberger, and F. Widmer, A new multinozzle encapsulation/immobilisation system to produce uniform beads of alginate, \textit{J. Biotechnol.}, 63, pp. 73-80 (1998).

\bibitem{ref:Haeberle2008}S. Haeberle, L. Naegele, R. Burger, F. von Stetten, R. Zengerle, and J. Ducrèe, Alginate bead fabrication and encapsulation of living cells under centrifugally induced artificial gravity conditions., \textit{J. Microencapsul.}, 25, pp. 267-274 (2008).

\bibitem{ref:Serp2000}D. Serp, E. Cantana, C. Heinzen, U. von Stockar, and I.W. Marison, Characterization of an encapsulation device for the production of monodisperse alginate beads for cell immobilization, \textit{Biotechnol. Bioengng.}, 70, pp. 41-53 (2000).

\bibitem{ref:Martins2014}I.M. Martins, M.F. Barreiro, M. Coelho, and A.E. Rodrigues, Microencapsulation of essential oils with biodegradable polymeric carriers for cosmetic applications, \textit{Biotechnol. Bioengng.}, 245, pp. 191 - 200 (2014).

\bibitem{ref:Augustin2009}M.A. Augustin, and Y. Hemar, Nano- and micro-structured assemblies for encapsulation of food ingredients, \textit{Chem. Soc. Rev.}, 38, pp. 902-912(2009).

\bibitem{ref:Riche2016}C.T. Riche, E.J. Roberts, M. Gupta, R.L. Brutchey, and N. Malmstadt, Flow invariant droplet formation for stable parallel microreactors, \textit{Nat. Commun.}, 7, 10780 (2016).

\bibitem{ref:Koester2008}S. Köster, F.E. Angilè, H. Duan, J.J. Agresti, A. Wintner, C. Schmitz, A.C. Rowat, C.A. Merten, D. Pisignano, A.D. Griffiths, and D.A. Weitz, Drop-based microfluidic devices for encapsulation of single cells, \textit{Lab Chip}, 8, pp. 1110-1115 (2008).

\bibitem{ref:Chabert2008}M. Chabert, and J.-L. Viovy, Microfluidic high-throughput encapsulation and hydrodynamic self-sorting of single cells, \textit{Proc. Natl. Acad. Sci. U.S.A.}, 105, pp. 3191-3196 (2008).

\bibitem{ref:Zhao2013}C.-X. Zhao, Multiphase flow microfluidics for the production of single or multiple emulsions for drug delivery, \textit{Adv. Drug Deliv. Rev.}, 65, 11, pp. 1420 - 1446 (2013).

\bibitem{ref:Chiu2017}D.T. Chiu, A.J. deMello, D. Di Carlo, P. S. Doyle, C. Hansen, R.M. Maceiczyk, and R.C.R. Wootton, Small but perfectly formed? Successes, challenges, and opportunities for microfluidics in the chemical and biological sciences, \textit{Chem}, 2, pp. 201 - 223 (2017).

\bibitem{ref:Dressaire2017}E. Dressaire, and A. Sauret, Clogging of microfluidic systems, \textit{Soft Matter}, 13, pp. 37-48 (2017).

\bibitem{ref:Su2006}F. Su, K. Chakrabarty, and R.B. Fair, Microfluidics-based biochips: Technology issues, implementation platforms, and design-automation challenges, \textit{IEEE Trans. Comput.-Aided Des. Integr. Circuits Syst}, 25, pp. 211-223 (2006).

\bibitem{ref:Sanchez2016}L. Duque Sánchez, N. Brack, A. Postma, P.J. Pigram, and L. Meagher, Surface modification of electrospun fibres for biomedical applications: A focus on radical polymerization methods, \textit{Biomaterials}, 106, pp. 24-45 (2016).

\bibitem{ref:Hu2014}X. Hu, S. Liu, G. Zhou, Y. Huang, Z. Xie, and X. Jing, Electrospinning of polymeric nanofibers for drug delivery applications, \textit{J Control Release}, 185, pp. 12-21 (2014).

\bibitem{ref:Enizi2018}A.M. Al-Enizi, M.M. Zagho, and A.A. Elzatahry, Polymer-based electrospun nanofibers for biomedical applications, \textit{Nanomaterials}, 259, 8 (2018).

\bibitem{ref:Moghe2008}A.K. Moghe, and B.S. Gupta, Co-axial electrospinning for nanofiber structures: Preparation and applications, \textit{Polym. Rev.}, 48, pp. 353-377 (2008).

\bibitem{ref:Yarin2011}A.L. Yarin, Coaxial electrospinning and emulsion electrospinning of core–shell fibers, \textit{Polym. Adv. Technol}, 22, pp. 310 - 317 (2011).

\bibitem{ref:Hu2006}H. Qi, P. Hu, J. Xu, and A. Wang, Encapsulation of drug reservoirs in fibers by emulsion electrospinning{:} Morphology characterization and preliminary release assessment, \textit{Biomacromolecules}, 7, 8, pp. 2327 - 2330 (2006).

\bibitem{ref:Yarin1995}A. Yarin, and D. Weiss, Impact of drops on solid surfaces: Self-similar capillary waves, and splashing as a new type of kinematic discontinuity, \textit{J. Fluid Mech.}, 283, pp. 141-173.

\bibitem{ref:Yonemoto2017}Y. Yonemoto, and T. Kunugi, Analytical consideration of liquid droplet impingement on solid surfaces, \textit{Sci. Rep.}, 7, 2362 (2017).

\bibitem{ref:Wildeman2016}S. Wildeman, C.W. Visser, C. Sun, and D. Lohse, On the spreading of impacting drops, \textit{J. Fluid Mech.}, 805, pp. 636-655 (2016).

\bibitem{ref:Eggers2010}J. Eggers, M.A. Fontelos, C. Josserand, and S. Zaleski, Drop dynamics after impact on a solid wall: Theory and simulations, \textit{Phys. Fluids}, 22, 062101 (2010).

\bibitem{ref:Laan2014}N. Laan, K.G. de Bruin, D. Bartolo, C. Josserand, and D. Bonn, Maximum diameter of impacting liquid droplets, \textit{Phys. Rev. Applied}, 2, 044018 (2014).

\bibitem{ref:Clanet2004}C. Clanet, C. Béguin, D. Richard, and D. Quéré, Maximal deformation of an impacting drop, \textit{J. Fluid Mech.}, 517, pp. 199-208 (2004).

\bibitem{ref:Josserand2003}C. Josserand, and S. Zaleski, Droplet splashing on a thin liquid film, \textit{Phys. Fluids}, 15, pp. 1650-1657 (2003).

\bibitem{ref:Wang2000} A.-B. Wang, and C.-C. Chen, Splashing impact of a single drop onto very thin liquid films, \textit{Phys. Fluids}, 12, pp. 2155-2158 (2000).

\bibitem{ref:Kittel2018}H.M. Kittel, I.V. Roisman, and C. Tropea, Splash of a drop impacting onto a solid substrate wetted by a thin film of another liquid, \textit{Phys. Rev. Fluids}, 3, 073601 (2018).

\bibitem{ref:Ray2015}B. Ray, G. Biswas, and A. Sharma, Regimes during liquid drop impact on a liquid pool, \textit{J. Fluid Mech.}, 768, pp. 492-523 (2015).

\bibitem{ref:Lhuissier2013}H. Lhuissier, C. Sun, A. Prosperetti, and D. Lohse, Drop fragmentation at impact onto a bath of an immiscible liquid, \textit{Phys. Rev. Lett.}, 110, 264503 (2013).

\bibitem{ref:Moqaddam2016} A.M. Moqaddam, S.S. Chikatamarla, and I. V. Karlin, Simulation of binary droplet collisions with the entropic lattice Boltzmann method, \textit{Phys. Fluids}, 28, 022106 (2016).

\bibitem{ref:Planchette2017}C. Planchette, H. Hinterbichler, M. Liu, D. Bothe, and G. Brenn, Colliding drops as coalescing and fragmenting liquid springs, \textit{J. Fluid Mech.}, 814, pp. 277-300 (2017).

\bibitem{ref:Roisman2012}I.V. Roisman, C. Planchette, E. Lorenceau, and G. Brenn, Binary collisions of drops of immiscible liquids, \textit{J. Fluid Mech.}, 690, pp. 512-535 (2012).

\bibitem{ref:Roisman2004}I.V. Roisman, Dynamics of inertia dominated binary drop collisions, \textit{Phys. Fluids}, 16, pp. 3438-3449 (2004).

\bibitem{ref:Ashgriz1990}N. Ashgriz , and J. Y. Poo, Coalescence and separation in binary collisions of liquid drops, \textit{J. Fluid Mech.}, 221, pp. 183-204 (1990).

\bibitem{ref:Planchette2018}C. Planchette, S. Petit, H. Hinterbichler, and G. Brenn, Collisions of drops with an immiscible liquid jet, \textit{Phys. Rev. Fluids}, 3, 093603 (2018).

\bibitem{ref:Kamperman2018}T. Kamperman, V.D. Trikalitis, M. Karperien, C.W. Visser, and J. Leijten, Ultrahigh-throughput production of monodisperse and multifunctional Janus microparticles using in-air microfluidics, \textit{ACS Appl. Mater. Interfaces}, 10, pp. 23433-23438 (2018).

\bibitem{ref:Visser2018}C.W. Visser, T. Kamperman, L.P. Karbaat, D. Lohse, and M. Karperien, In-air microfluidics enables rapid fabrication of emulsions, suspensions, and 3D modular (bio)materials, \textit{Sci. Adv.}, 4, 1 (2018).

\bibitem{ref:Hinterbichler2015}H. Hinterbichler, C. Planchette, and G. Brenn, Ternary drop collisions, \textit{Exp. Fluids}, 56, 190 (2015).

\bibitem{ref:Planchette2012}C. Planchette, E. Lorenceau, and G. Brenn, The onset of fragmentation in binary liquid drop collisions, \textit{J. Fluid Mech.}, 702, pp. 5-25 (2012).

\bibitem{ref:Chen2006}R.-H. Chen, S.-L. Chiu, and T.-H. Lin, Collisions of a string of water drops on a water jet of equal diameter, \textit{Exp. Therm. Fluid Sci.}, 31, pp. 75-81 (2006).

\bibitem{ref:Baumgartner2020}D. Baumgartner, R. Bernard, B. Weigand, G. Lamanna, G. Brenn, and C. Planchette, Influence of liquid miscibility and wettability on the structures produced by drop-jet collisions, \textit{J. Fluid Mech.}, 885, A23 (2020).

\bibitem{ref:Brenn1996}G. Brenn, F. Durst, and C. Tropea, Monodisperse sprays for various purposes - their production and characteristics, \textit{Part. Part. Syst. Charact.}, 13, 179-185 (1996).

\bibitem{ref:Peters2013}F. Peters, and D. Arabali, Interfacial tension between oil and water measured with a modified contour method, \textit{Colloids and Surfaces A}, 426, pp. 1-5 (2013). 
 
\bibitem{ref:Robinson1970}D. Robinson, and S. Hartland, The shape of liquid drops approaching a deformable liquid—liquid interface in three-phase systems, \textit{Chem. Eng. J.}, 1, pp. 22-30 (1970). 

\bibitem{ref:Dong2017}T. Dong, W.H. Weheliye, P. Chausset, and P. Angeli, An experimental study on the drop/interface partial coalescence with surfactants, \textit{Phys. Fluids}, 29, 102101 (2017). 

\bibitem{ref:Baumgartner2019}D. Baumgartner, P. Benez, G. Brenn, and C. Planchette,  Liquid/Liquid encapsulation: effects of wettability and miscibility, 29th Conference on Liquid Atomization and Spray Systems, Paris, France, Sep. 2-4 (2019).

\bibitem{ref:Planchette2017ILASS}C. Planchette, H. Hinterbichler, and G. Brenn,  Drop stream - Immiscible jet collisions{:} Regimes and fragmentation mechanisms, Valencia, Spain, Sep. 6-8 (2017).

\bibitem{ref:Rayleigh1892}Lord Rayleigh Sec. R.S., XVI. On the instability of a cylinder of viscous liquid under capillary force, \textit{The London, Edinburgh, and Dublin Philosophical Magazine and Journal of Science}, 207, pp. 145-154 (1892).

\bibitem{ref:Shlegel2020}N.E. Shlegel, P.P. Tkachenko, and P.A. Strizhak, Influence of viscosity, surface and interfacial tensions on the liquid droplet collisions, \textit{Chem. Eng. Sci.}, 220, 115639 (2020).

\bibitem{ref:Walzel1980}P. Walzel, Zerteilgrenze beim Tropfenprall, \textit{Chem.-Ing.-Tech.}, 52, pp. 338-339 (1980).

\bibitem{ref:Geppert2017}A. Geppert, A. Terzis, G. Lamanna, M. Marengo, and B. Weigand, A benchmark study for the crown-type splashing dynamics of one- and two-component droplet wall–film interactions, \textit{Exp. Fluids}, 58, 172 (2017). 

\bibitem{ref:Stone1986}H.A. Stone, B.J. Bentley, and L.G. Leal, An experimental study of transient effects in the breakup of viscous drops, \textit{J. Fluid Mech.}, 173, pp. 131-158 (1986).

\bibitem{ref:Stone1989}H.A. Stone, and L.G. Leal, Relaxation and breakup of an initially extended drop in an otherwise quiescent fluid, \textit{J. Fluid Mech.}, 198, pp. 399-427 (1989).

\bibitem{ref:Stone1994}H.A. Stone, Dynamics of drop deformation and breakup in viscous fluids, \textit{Annu. Rev. Fluid Mech.}, 26, pp. 65-102 (1994).

\bibitem{ref:Grace1982}H.P. Grace, Dispersion phenomena in high viscosity immiscible fluid systems and application of static mixers as dispersion devices in such systems, \textit{Chem. Eng. Commun.}, 14, pp. 225–277 (1982).

\bibitem{ref:supplmat}See Supplemental Material at [URL] for the evolution of the aspect ratio $\zeta = L_{max}/D_d$ with $We_d$ and the jet deformation. 

\bibitem{ref:Richard2002}D. Richard, C. Clanet, and D. Quéré, Contact time of a bouncing drop, \textit{Nature}, 417, 811 (2002).

\bibitem{ref:Bakshi2007}S. Bakshi, I.V. Roisman, and C. Tropea, Investigations on the impact of a drop onto a small spherical target, \textit{Phys. Fluids}, 19, 032102 (2007).

\bibitem{ref:Huang2018}H.-M. Huang, and X.-P. Chen, Energetic analysis of drop’s maximum spreading on solid surface with low impact speed, \textit{Phys. Fluids}, 30, 022106 (2018).


\bibitem{ref:Roisman2009a}I.V. Roisman, E. Berberović, and C. Tropea, Inertia dominated drop collisions. I. On the universal flow in the lamella, \textit{Phys. Fluids}, 21, 052103 (2009).

\bibitem{ref:Roisman2009b}I.V. Roisman, Inertia dominated drop collisions. II. An analytical solution of the Navier–Stokes equations for a spreading viscous film, \textit{Phys. Fluids}, 21, 052104 (2009).


\bibitem{ref:Rioboo2002}R. Rioboo, M. Marengo, and C. Tropea, Time evolution of liquid drop impact onto solid, dry surfaces, \textit{Exp. Fluids}, 33, pp. 112-124 (2002).



\end{thebibliography}
\end{document}